\documentclass[12pt]{article}
\usepackage[T1]{fontenc}
\usepackage{baskervald}
\usepackage[font=small, labelfont=bf]{caption}
\usepackage[labelformat=simple]{subcaption}

\usepackage{graphicx}
\usepackage{hyperref}
\usepackage{amsmath}
\usepackage{amssymb}
\usepackage{amsthm}
\usepackage{textcomp,bbm}
\usepackage{longtable}
\usepackage{verbatim}
\usepackage{cite}
\usepackage{enumerate}

\newtheoremstyle{named}{}{}{\itshape}{}{\bfseries}{.}{.5em}{\thmnote{#3 }#1}
\theoremstyle{named}
\newtheorem*{namedconjecture}{Conjecture}

\usepackage[usenames,dvipsnames]{xcolor}

\theoremstyle{definition}

\newcommand{\be}{\begin{equation}}
\newcommand{\ee}{\end{equation}}
\newcommand{\ba}{\begin{align}}
\newcommand{\ea}{\end{align}}
\newcommand{\beq}{\begin{eqnarray}}
\newcommand{\eeq}{\end{eqnarray}}

\def\simlt{\stackrel{<}{{}_\sim}}

\usepackage{titlesec}

\begin{document}

\title{Axion Experiments to Algebraic Geometry\\ --- \\ \Large Testing Quantum Gravity\\ via the Weak Gravity Conjecture}

\author{Ben Heidenreich,\thanks{Corresponding author}\ \ Matthew Reece, and Tom Rudelius\\
{\small \texttt{bjheiden, mreece, rudelius~(@physics.harvard.edu)}}\\
Department of Physics, Harvard University\\
Jefferson Laboratory, 17 Oxford St., Cambridge, MA, 02138}

\date{\today}

\maketitle

\begin{abstract}
Common features of known quantum gravity theories may hint at the general nature of quantum gravity. The absence of continuous global symmetries is one such feature. This inspired the Weak Gravity Conjecture, which bounds masses of charged particles. We propose the Lattice Weak Gravity Conjecture, which further requires the existence of an infinite tower of particles of all possible charges under both abelian and nonabelian gauge groups and directly implies a cutoff for quantum field theory. It holds in a wide variety of string theory examples and has testable consequences for the real world and for pure mathematics. We sketch some implications of these ideas for models of inflation, for the QCD axion (and LIGO), for conformal field theory, and for algebraic geometry.\footnote{Essay written for the Gravity Research Foundation 2016 Awards for Essays on Gravitation.}
\end{abstract}

\newpage

\section{Introduction} 

Quantum gravity is famously difficult. Its effects are tiny at accessible energies, whereas our theoretical understanding of it relies largely on supersymmetry, which is notably absent from the real world (at least at accessible energies). Making predictions about our universe, rather than a pristine toy model, is challenging. For this reason, any feature shared by all known consistent quantum gravity theories could be an important clue to their general nature.

We recently identified a candidate for such a feature: the Lattice Weak Gravity Conjecture (LWGC) \cite{Heidenreich:2015nta}. A powerful statement about the nature of quantum gravity, it passes stringent theoretical checks and makes nontrivial predictions for real-world observable physics, properties of as-yet-unformulated theories of quantum gravity, and pure mathematics. In this essay we motivate and provide evidence for the LWGC (the many well-understood examples at least call for a deeper explanation) and explain some of its implications. A virtue of this conjecture is that it points the way toward many tractable theoretical questions whose answers
 will provide more evidence for the conjecture or shed light on the nature of quantum gravity theories in which it fails.

The Weak Gravity Conjecture (WGC) \cite{ArkaniHamed:2006dz} makes quantitative the qualitative belief that theories of quantum gravity should have no continuous global symmetries \cite{Banks:1988yz,Kallosh:1995hi,Susskind:1995da,Banks:2010zn}: as Hawking radiation is unaffected by global symmetry charges, this would lead to an infinite number of stable black hole remnants near the Planck scale, with associated thermodynamic problems. A gauge theory at extremely weak coupling $e \to 0$ is nearly indistinguishable from a global symmetry, so we expect quantum gravity theories to resist the $e \to 0$ limit.

In the real world charged black holes will discharge by emitting electrons. This observation was elevated to a postulate by \cite{ArkaniHamed:2006dz}: in the absence of special BPS conditions forbidding decay, all charged black holes should spontaneously discharge. Charged black holes in general relativity have a maximum charge for a given mass. For the decay of maximally-charged black holes to be kinematically possible, a charged particle must exist with:
\begin{equation}
m \leq \sqrt{2} e q M_{\rm Planck}.
\end{equation}
A further argument in \cite{ArkaniHamed:2006dz} based on magnetic monopoles suggested the theory has a cutoff length scale $(e M_{\rm Planck})^{-1}$ and hence breaks down as $e \to 0$.

\begin{figure}
  \centering
  \begin{subfigure}[b]{0.4\textwidth}
  \centering
  \includegraphics[width=\textwidth]{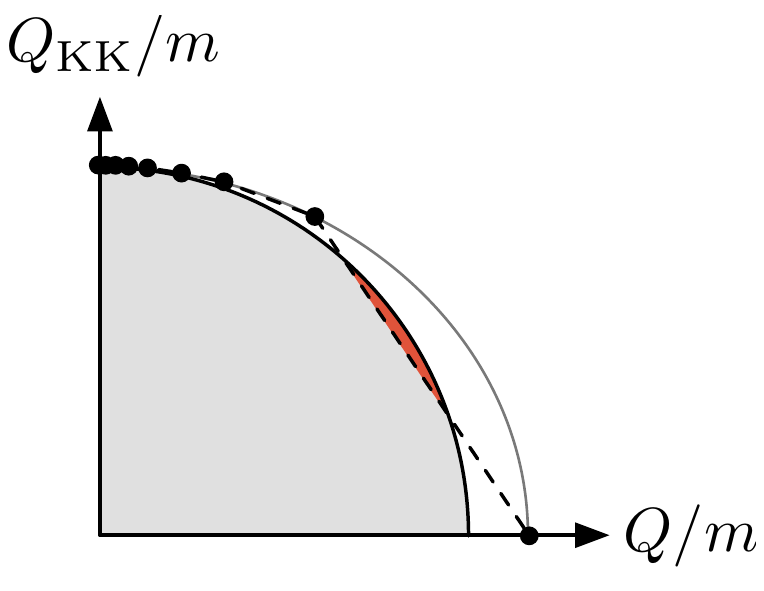}
  \caption{WGC violation after compactification} \label{sfig:KKviolation}
  \end{subfigure}
  \hfill
  \begin{subfigure}[b]{0.4\textwidth}
  \centering
  \includegraphics[width=\textwidth]{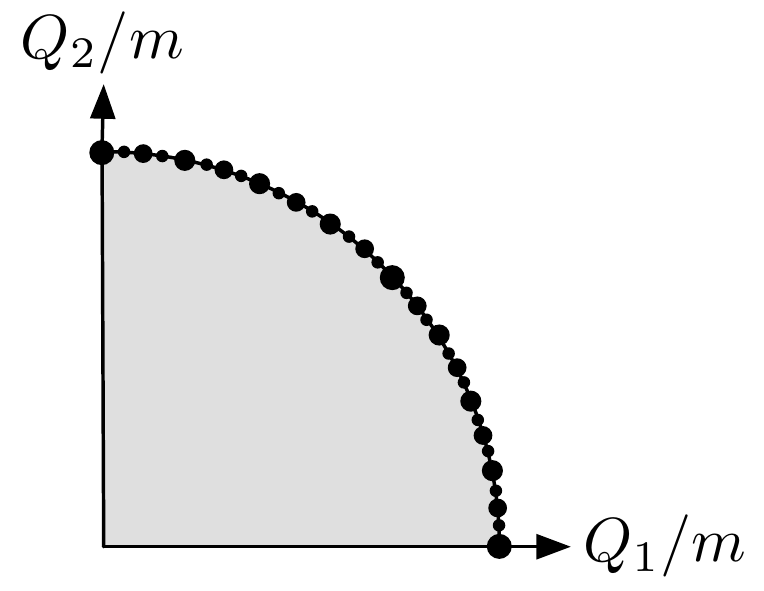}
  \caption{The LWGC solution} \label{sfig:T2KK}
  \end{subfigure}
  \caption{\subref{sfig:KKviolation}~Kaluza-Klein reduction of theories satisfying the WGC can lead to violations in the lower dimensional theory. Sub-extremal black holes with charge-to-mass ratios in the red region are stable. \subref{sfig:T2KK}~If the higher dimensional photon is itself a Kaluza-Klein photon, then there is an extremal charged particle of every charge, and the WGC is satisfied.} \label{fig:KKinvestigation}
\end{figure}

Our refined WGC gives a different perspective on the $e \to 0$ limit:
\begin{namedconjecture}[Lattice Weak Gravity]
For any charge $\vec{Q}$ allowed by charge quantization, a particle of charge $\vec{Q}$ exists with charge-to-mass ratio at least as large as that of a semiclassical extremal black hole with $\vec{Q}_{\rm BH} \propto \vec{Q}$.
\end{namedconjecture}
\noindent Here $\vec{Q}$ is a multicomponent vector in theories with multiple massless photons, and refers to the Cartan charges when there are non-abelian gauge symmetries. This statement is much stronger than the original WGC as it requires the existence of infinitely many charged particles (though most are effectively black holes). This is not a step we take lightly. We were led to this hypothesis by investigating Kaluza-Klein theories, as shown in Figure~\ref{fig:KKinvestigation}.

The LWGC demands one particle lighter than $e M_{\rm Pl}$, another lighter than $2 e M_{\rm Pl}$, another $3 e M_{\rm Pl}$, and so on: an infinite tower of charged states. Quantum field theory always breaks down in the presence of such a tower, giving a completely different explanation for the cutoff length scale $(e M_{\rm Pl})^{-1}$ from that of \cite{ArkaniHamed:2006dz}. With the LWGC, infinitely many particles become massless when $e \to 0$, implying the theory breaks down completely in the limit of continuous global symmetries, as expected.

We speculate that the LWGC is required for a completely consistent picture of black hole thermodynamics \cite{Preskill:1991tb} (holography \cite{Harlow:2015lma} or Cosmic Censorship \cite{Horowitz:2016ezu} may also lead to general arguments), but at present the primary evidence for the conjecture comes from the wide variety of well-understood quantum gravity theories where it holds. We have checked it in perturbative heterotic string theory, including toroidal compactifications with Wilson lines; supersymmetric compactifications on $\mathbb{Z}_N$ orbifolds of $T^4$ or $T^6$ without Wilson lines; certain examples of Higgs branches; and some $T^8$ orbifolds lacking one-loop tadpoles. The only examples we know where it appears to fail are nonsupersymmetric theories with instabilities, for which our perturbative calculations are unreliable.\footnote{Note added: since the initial submission of this essay to the Gravity Research Foundation we have found counterexamples in certain supersymmetric orbifold compactifications of the heterotic string to the precise formulation of the Lattice WGC presented here. However, we are gathering evidence that a closely related variant of it is consistent with a wide class of string theory examples in which we have computed the full spectrum. We will present the full details in a forthcoming paper.} The LWGC also predicts that finite size black holes can be slightly superextremal, as verified in some string examples by~\cite{Kats:2006xp}. Thus we have a vast (and growing) set of examples of quantum gravity theories obeying the LWGC.

\section{Applications}

The WGC connects to a remarkable array of topics in high energy physics, of which we highlight a few: axion inflation, the QCD axion, AdS/CFT, and Calabi-Yau geometry.

\subsection{Axion Inflation}

The best-developed application of the WGC is to large-field axion inflation, which predicts detectable primordial gravitational waves. Viewing axions as 0-form analogues of gauge fields, the WGC generalizes to the existence of an instanton of action $S$ satisfying
\begin{equation}
S \simlt M_{\rm Planck}/f. \label{eq:WGCaxion}
\end{equation}
Axions in controlled regimes of string theory obey this bound \cite{Banks:2003sx}. String theory requires $S \gtrsim 1$ to maintain perturbative control while successful inflation requires $f \gtrsim 5 M_{\rm Planck}$; the WGC implies these are incompatible. 

\begin{figure}
\begin{center}
\includegraphics[trim=15mm 10mm 15mm 10mm, clip, width=68mm]{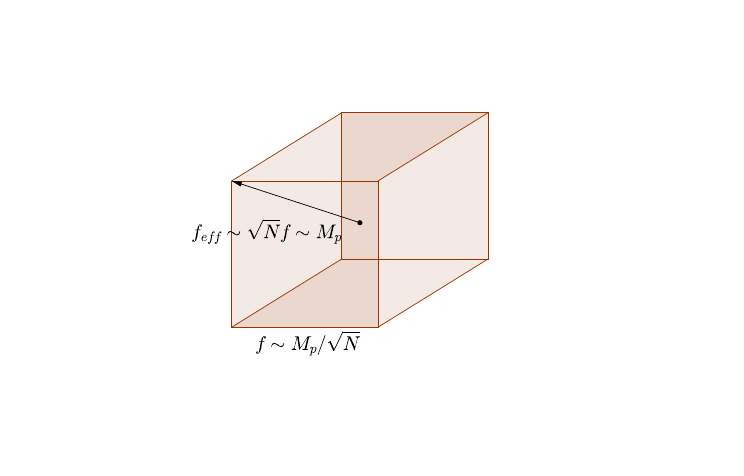}
\includegraphics[trim=15mm 10mm 15mm 10mm, clip, width=68mm]{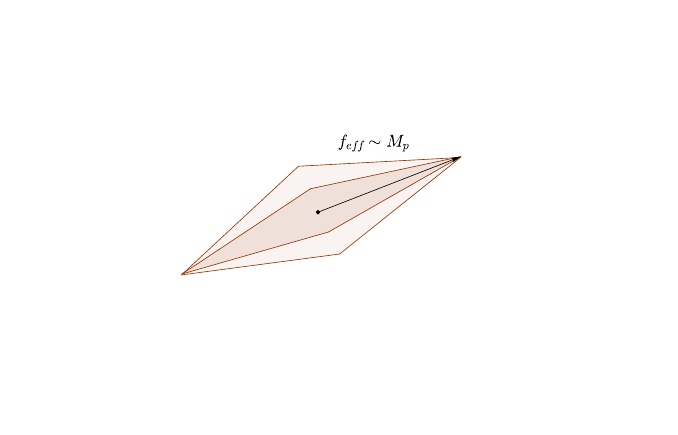}
\caption{N-flation (left) and decay constant alignment (right) seek to produce a large effective decay constant from multiple sub-Planckian ones.  However, the WGC implies $f_{\rm eff} \lesssim M_{\rm Planck}$ for the simplest versions of these models.}
\label{Nandalign}
\end{center}
\end{figure}

Models evading this bound posit many axions each traversing a smaller field range during inflation (``N-flation'' \cite{Liddle:1998jc,Copeland:1999cs, Dimopoulos:2005ac}) or at least two axions, with the inflaton taking a long diagonal path within a small field space (``alignment'' \cite{Kim:2004rp}), as in Figure \ref{Nandalign}. Recent arguments cast doubt on these models \cite{Rudelius:2015xta, Montero:2015ofa,Brown:2015iha, Heidenreich:2015wga} using a generalized WGC for theories of multiple gauge fields \cite{Cheung:2014vva}. For N-flation as well as many alignment models, the total inflaton displacement is bounded to be sub-Planckian.

Large field natural inflation models consistent with the WGC are being actively explored \cite{delaFuente:2014aca,Hebecker:2015rya, Bachlechner:2015qja, Kappl:2015esy}, but may not exist as consistent quantum gravity theories \cite{Rudelius:2014wla, Junghans:2015hba, Conlon:2016aea, Baume:2016psm, Long:2016jvd}. The WGC's consequences for axion monodromy \cite{Silverstein:2008sg,McAllister:2008hb} and the relaxion \cite{Graham:2015cka} are under intense investigation\cite{Heidenreich:2015wga,Ibanez:2015fcv,Hebecker:2015zss}.

\subsection{QCD Axion}

A major particle physics puzzle is the tiny QCD theta angle: ${\bar \theta} \simlt 10^{-10}$. The most plausible solution is a light pseudoscalar field, the QCD axion, which dynamically relaxes $\bar \theta$ \cite{Kim:2008hd}. The QCD instanton action is
\beq
S_{\rm QCD} = 4 \ln{M_*/\Lambda_{\rm QCD}} \approx 160 \,,
\eeq
where $M_*$ is an ultraviolet scale such as $M_{\rm GUT}$. Using the general bound (\ref{eq:WGCaxion}) we predict
\beq
f_{\rm QCD} \lesssim 10^{16}~{\rm GeV} \,,
\eeq
with mild logarithmic uncertainties. Experimental observation of a QCD axion with decay constant well above the GUT scale would contradict the WGC.

A common claim is that early-universe cosmology requires $f_{\rm QCD} < 10^{12}~{\rm GeV}$ to avoid dark matter overproduction, but alternative cosmologies \cite{Kawasaki:1995vt,Banks:2002sd,Acharya:2010zx} or initial conditions \cite{Linde:1991km,Wilczek:2004cr} allow larger $f_{\rm QCD}$. The absence of QCD axions with decay constants above the GUT scale is a nontrivial, falsifiable prediction of the WGC. It can be tested in terrestrial searches for axion dark matter \cite{Graham:2011qk,Budker:2013hfa,Kahn:2016aff}. Axions with decay constants above the GUT scale also cause superradiant instabilities in black holes, detectable through astrophysical observations including gravitational wave signals \cite{Arvanitaki:2009fg, Arvanitaki:2010sy, Arvanitaki:2014wva,Arvanitaki:2016qwi}. The new window on gravitational physics that LIGO has opened can also test the WGC!

\subsection{AdS/CFT}

The AdS/CFT correspondence encompasses many quantum gravity theories. The simplest 4d CFT translation of the WGC in AdS$_5$ is the existence of a charge $Q$, dimension $\Delta$ operator with
\begin{equation} \label{eqn:WGCAdS}
\frac{\Delta}{\sqrt{12 c}} \leq \frac{Q}{\sqrt{b}},
\end{equation}
with $c$ the central charge and $b$ the coefficient of the current two-point function. However, the flat-space WGC may be modified due to the curvature of AdS \cite{Nakayama:2015hga} (cf.~\cite{Benjamin:new} on AdS$_3$/CFT$_2$). We have checked that the naive analog of the LWGC based on~(\ref{eqn:WGCAdS}) holds for ${\cal N}=4$ super-Yang-Mills theory. Further exploration of the LWGC in conformal field theory should be rewarding.

\subsection{Geometry}

In some string theory compactifications, the masses and charges of particles or branes are related to  geometrical properties of the compactification manifold. This means that the LWGC, if true, has corollaries in pure mathematics. (A related discussion of the ``geometric weak gravity conjecture'' appears in \cite{Hebecker:2015zss}, though they have not considered the implications of the stronger Lattice WGC.) Compactifying M-theory on a Calabi-Yau three-fold gives a 5d supergravity theory with BPS particle states corresponding to M2-branes wrapping holomorphic curves in the three-fold. These BPS states generate the ``effective'' cone in the charge lattice, inside which the BPS bound requires $|\vec{Q}| \leq M$ in Planck units. Simultaneously satisfying this bound as well as the LWGC requires a BPS state of $|\vec{Q}|=M$ at every lattice point within the cone. Geometrically, this means every effective 2-cycle class of a compact Calabi-Yau should contain a holomorphic curve representative. This conjecture is similar in spirit to Vafa's inference of the existence of an exponentially large number of holomorphic curves in 2-cycle classes far out in the homology lattice \cite{Vafa:1997gr} in order to match microscopic BPS degeneracies with black hole entropy \cite{Strominger:1996sh}. The conjecture is quite surprising from a mathematical perspective, though it has been verified in the simple case of the quintic three-fold \cite{Clemens, Katz86} and Sen \cite{Sen:1995vr} has proven it for $T^6$ using U-duality \cite{Hull:1994ys}.

\subsection{The Nature of Quantum Gravity}

The LWGC holds in a wide variety of perturbative string vacua and beyond. It encapsulates a central feature of string theory: the existence of a tower of increasingly heavy states of ascending charge. We imagine that a further refinement of the WGC might capture even more of the essence of string theory. For instance, it is tempting to postulate that states of arbitrary {\em spin} exist at any point in the charge lattice, with masses gradually increasing for higher spins. The challenge is to formulate such conjectures in a precise manner, consistent with all available theoretical data. Our dream is that by examining many individual consistent realizations of quantum gravity provided by string theory, we can abstract away the details of specific vacua and gain a clear, testable vision of the essential nature of quantum gravity. 

\section*{Acknowledgments}
We thank Cumrun Vafa and Shing-Tung Yau for useful discussions. BH is supported by the Fundamental Laws Initiative of the Harvard Center for the Fundamental Laws of Nature. The work of MR is supported in part by the NSF Grant PHY-1415548. TR is supported in part by the National Science Foundation under Grant No.~DGE-1144152. 

\bibliography{essayref}
\bibliographystyle{utphys}

\end{document}